\documentclass[twocolumn,english,showkeys,superscriptaddress,longbibliography]{revtex4-1}
\usepackage[utf8x]{inputenc}
\usepackage{amsfonts}
\usepackage{amsmath}
\usepackage{amssymb}
\usepackage[colorlinks, citecolor=blue,linkcolor=red]{hyperref}
\usepackage{color}
\usepackage{float}
\usepackage{hyperref}
\usepackage{lipsum}
\usepackage{textcomp}
\usepackage[rightcaption]{sidecap}
\usepackage{subfigure}
\usepackage[rightcaption]{sidecap}
\usepackage{graphicx}
\begin{document}
\title{Multi-Stability in Cavity QED with Spin-Orbit Coupled Bose-Einstein Condensate}
\author{Kashif Ammar Yasir}
\email{kayasir@zjnu.edu.cn}\affiliation{Department of Physics, Zhejiang Normal University, Jinhua 321004, China.}
\author{Gao Xianlong}
\email{gaoxl@zjnu.edu.cn}
\affiliation{Department of Physics, Zhejiang Normal University, Jinhua 321004, China.}
\setlength{\parskip}{0pt}
\setlength{\belowcaptionskip}{-10pt}
\begin{abstract}
We investigate the occurrence of steady-state multi-stability in a cavity system containing spin-orbit coupled Bose-Einstein condensate and driven by a strong pump laser. The applied magnetic field splits the Bose-Einstein condensate into pseudo-spin states, which then became momentum sensitive with two counter propagating Raman lasers directly interacting with ultra-cold atoms. After governing the steady-state dynamics for all associated subsystems, we show the emergence of multi-stable behavior of cavity photon number, which is unlike with previous investigation on cavity-atom systems. However, this multi-stability can be tuned with associated system parameters. Further, we illustrate the occurrence of mixed-stability behavior for atomic population of the pseudo spin-$\uparrow$ amd spin-$\downarrow$ states, which are appearing in so-called bi-unstable form. The collective behavior of these atomic number states interestingly possesses a transitional interface among the population of both spin states, which can be enhance and controlled by spin-orbit coupling and Zeeman field effects. Furthermore, we illustrate the emergence of secondary interface mediated by increasing the mechanical dissipation rate of the pseudo-spin states. These interfaces could be cause by the non-trivial behavior of synthetic spin state mediated by cavity. Our findings are not only crucial for the subject of optical switching, but also could provide foundation for future studies on mechanical aspect of synthetic atomic states with cavity quantum electrodynamics.         
   
\end{abstract}
\date{\today}
\maketitle

\section{Introduction}
Cavity quantum electrodynamics (QED) -- strong photonic modes confined in a highly reflective oppositely facing mirrors -- is appeared to be the fascinating tool to generate coherent photonic interaction \cite{Ref1,Ref2,Ref3,Ref4}. The interaction of these confined and coherent optical modes with other physical objects, like atoms and ultra-cold atoms, further enhances the advantages of cavity QED \cite{Ref5,Ref6,Ref7}. This combination of coherent atom-light interaction has the ability to even practically test the foundation of quantum computation \cite{Ref8,Ref9}. Further, the inclusion of cold and ultra-cold atoms in the cavity QED (specially in cavity-optomechanics where the mechanical effects of light manipulates the motion of one end mirror(s)), provided another direction of multi-species (hybrid) cavity QED \cite{Esslinger}. The hybrid cavity QED led to the study of Hamiltonian chaos \cite{Meystre2010}, chaos induced quantum mechanical localization \cite{kashif1,kashif2}, atom induced optomechanical cooling \cite{peter2}, multi-particle quantum entanglement \cite{Vitali2012,Vitali2014,Sete2014,Hofer} and high-fidelity state transfer \cite{YingPRL2012,SinghPRL2012}. Quantum nonlinear optics with hybrid cavity QED further results in the concept of multiple electromagnetically induced transparencies \cite{agarwal2010,Stefan2010,Peng2014,SafaviNaeini2011,kashif11,kashif111}. The manipulation of synthetic inter-atomic states of ultra-cold atoms in hybrid optomechanical systems lead to the further enhancements of hybrid cavity QED operations \cite{kashif3,kashif33,kashif4}.

Meanwhile, the spin-orbit (SO-) coupling, a crucial relation between the momentum of a quantum particle and its own spin \cite{Ref32,Ref33,Ref34}, is proven to be the key tool to study spin-Hall effect \cite{Ref35,Ref36} and topological insulators \cite{Ref37,Ref38,Ref39,Ref40}, and is the subject of increasing investigations \cite{Ref41,Ref42,Ref43,Ref44}. The recent demonstrations of SO-coupling Bose-Einstein condensate (BEC) in optical cavities \cite{Ref45,Ref46,Ref47} and optomechanical systems, where the SO-coupling has been used to cool the mechanical mirror \cite{Ref48}, have provided basis to manipulate photonic interactions at pseudo-spin state level. The SO-coupling mediated dressed states interactions in an optical cavity further led to the discovery of topological states in the transmitting probe light \cite{kashif55}. However, in spite of these investigation, still a study on steady-state stability behavior is desirable because it will lead to the dynamical aspect of such hybrid systems.  

In this paper, we demonstrated the opportunity to achieved multi-stability in the steady-state behavior of a cavity with SO-coupled BEC and driven by a single-mode external pump laser. The externally interacting magnetic field excites atomic pseudo-spin states and two counter propagating Raman beams directly interact with confined atomic modes in cavity resulting in SO-coupling between magnetically engineered pseudo-spin states. We compute steady-state behavior for all associated degrees of freedom from quantum Langevin equations driven from total Hamiltonian of the system. From governed steady-state equations, show that the intra-cavity photon number not only possesses multiple stable states (or multi-stable behavior) unlike previous studies, but it is also tunable with other associated system parameters. Further, we illustrate a unique and mixed stable and unstable behavior of steady-state atomic population in both spin states because of the Raman process mediated SO-coupling and Raman coupling. We also demonstrate the dependence of both populations on each other by observing their collective behavior versus external laser power. We found an intersection between these atomic numbers that could be because of the phase transition occurring in dress states with SO-coupling. This feature can be enhanced and controlled with associated parameters, especially with atomic mechanical damping.        

The manuscript is organized as: Section \ref{sec1} contains the system modeling and details of mathematical calculations. Section \ref{sec2} illustrates the results of steady-state behavior of intra-cavity photon number. While the section \ref{sec3} contains the results of multi-stability occurring in the steady-state behavior of pseudo-spin states. Finally, section \ref{sec4} contains the conclusion of the study. 

\section{System Description and Hamiltonian}\label{sec1}
\begin{figure}[tp]
	\includegraphics[width=6cm]{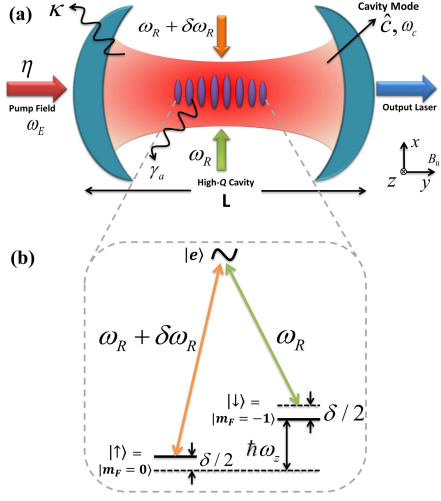}
	\caption{(a) The schematic diagram of a high-$Q$ Fabry-P\'erot cavity containing spin-orbit coupled Bose-Einstein condensate. A pump laser $\eta$ (with frequency $\omega_E$) drives the cavity exciting strong cavity mode, which then generates coupling between cavity and trapped spin-orbit coupled Bose-Einstein condensate. A bias magnetic field with strength $B_0$ interacts with the atomic states trapped along $\hat{y}$-axis causing Zeeman splitting $\hbar\omega_z$ between atomic pseudo states. While two counter propagating Raman lasers interact transversely along $\hat{x}$-axis with atoms to generate spin-orbit coupling. (b) The energy level excitation diagram illustrating the excitations induced by the bias magnetic field $B_0$ and counter propagating Raman lasers.}
	\label{fig1}
\end{figure}
The $N\approx1.8\times10^5$ $^{87}Rb$ bosonic particles are trapped in a high-$Q$ Fabry-P\'erot cavity having length $L\approx12.5\times10^{-3}$m, see Fig.\ref{fig1}(a) \cite{Refnew0,Refnew2,Refnew1,Ref49}. External pump laser with frequency $\omega_{E}$ and strength $\vert\eta\vert=\sqrt{P\times\kappa/\hbar\omega_{E}}$, where $P$ is the power, drives the cavity generating a strong cavity mode $\omega_{c}\approx1.9\times2\pi GHz$ having detuning $\Delta_c=\omega_E-\omega_c$ and decay rate $\kappa$.  
The $10 G$ of a bias magnetic field $B_0$ interacts with ultra-cold atoms trapped inside cavity and produces a Zeeman splitting $\hbar\omega_z$ under condition $|\omega_z/\kappa|>>1$ \cite{Ref45}.
After that, in order to generate SO-coupling, two counter propagating Raman lasers ($\omega_R$ and $\omega_R+\delta\omega_R$), having wavelength $\lambda=804.1 nm$ and detuning $\delta=1.6E_R$ \cite{Ref34}, transversely excite atoms along $\hat{x}$-axis. It yields in the coupling between two internal pseudo-spin states, within the spectrum of $5S_{1/2}$, of ultra-cold atoms ($|\uparrow\rangle =|F=2, m_F=0\rangle$ and $|\downarrow\rangle=|F=2, m_F=-1\rangle$) at electronic manifold $F=2$, see Fig. \ref{fig1}(b). We selected these particular parametric values from the recent and available experimental studies \cite{Refnew0,Refnew2,Refnew1} in literature to make our findings experimentally feasible. But our findings could be valid for a range of selected parametric values crucially depending upon the coupling strengths. 

The total Hamiltonian of the considered system, with adiabatic and rotating-wave approximation \cite{Ref34,Ref46,Ref47,Ref48}, reads as,
\begin{eqnarray}\label{Ha1}
	\hat{\mathcal{H}} &=& \int d\pmb{r}\pmb{\hat{\psi}^{\dag}}(\pmb{r})\bigg(\hat{\mathcal{H}}_{0}+\mathcal{V}_{Lat}\bigg) \pmb{\hat{\psi}}(\pmb{r})\nonumber\\
	&+&\frac{1}{2}\int d\pmb{r}\sum_{\sigma,\acute{\sigma}} \mathcal{U}_{\sigma,\acute{\sigma}}\hat{\psi}^{\dag}_\sigma(\pmb{r})\hat{\psi}^{\dag}_{\acute{\sigma}}(\pmb{r})\hat{\psi}_{\acute{\sigma}}(\pmb{r})\hat{\psi}_\sigma(\pmb{r})\nonumber\\
	&+&\hbar\Delta_{c}\hat{c}^{\dag}\hat{c}-i\hbar\eta(\hat{c}
	-\hat{c}^{\dag}).
\end{eqnarray}
Here $\pmb{\hat{\psi}}=[\hat{\psi}_\uparrow,\hat{\psi}_\downarrow]^{T}$ represents the bosonic field operators corresponding to atomic pseudo-spin states $|\uparrow\rangle$ and $|\downarrow\rangle$. 
$\hat{\mathcal{H}}_{0}=\hbar^2\pmb{k}^2\sigma_{0}/2m_a+\tilde{\alpha}k_x\sigma_{y}+\frac{\delta}{2}\sigma_y+\frac{\Omega_z}{2}\sigma_z$ 
is the single particle Hamiltonian accommodating spin-orbit interactions with strength $\tilde{\alpha}=E_R/k_L$. Here $\delta=-g\mu_B B_z$ is the Raman detuning and $\Omega_z=-g\mu_B B_y$ is the Raman coupling corresponding to the magnetic field effects along the $\hat{z}$ and $\hat{y}$-axis, respectively \cite{Ref34,Ref48}. The quasi-momentum is in one-dimension $\pmb{k}=[k_x,0,0]$ because of the one dimensional SO-coupling occurring along $\hat{x}$-axis. $\sigma_{x,y,z}$ represents Pauli matrices with unit matrix $\sigma_0$.
The two-dimensional lattice potential induced by the longitudinal (cavity mode driven by pump laser) and transverse (counter propagating Raman beans) fields reads as $\mathcal{V}_{Lat}=\hbar \hat{c}^{\dag}\hat{c}U_0[cos^2(kx)+cos^2(ky)]$ \cite{Ref48}, where $\hat{c}$ and $\hat{c}^\dag$ correspond to the annihilation and creation operator of cavity mode with wave number $k$, respectively. $U_0=g_0^2/\Delta_a$ corresponds to the potential depth defined by the Rabi oscillations $g_{0}$ and cavity-atomic detuning $\Delta_{a}$ \cite{Ref48}. The atom-atom interaction can be defined by $\mathcal{U}_{\sigma,\acute{\sigma}}=4\pi a_{\sigma,\acute{\sigma}}^2\hbar^2/m_a$, where $a_{\sigma,\acute{\sigma}}$ is the s-wave scattering. Last two terms correspond to the intra-cavity optical mode and its relation with pump drive, respectively, as mentioned before. 

We assume that the intra-species and inter-species interactions of pseudo-spin state as $U_{\uparrow,\uparrow}=U_{\downarrow,\downarrow}=U$ and $U_{\uparrow,\downarrow}=U_{\downarrow,\uparrow}=\varepsilon U$, respectively, with laser configuration parameter $\varepsilon$ \cite{Ref34,Ref48}.
After that, we insert plane-wave ansatz $\pmb{\hat{\psi}}(r)=e^{ikr}\pmb{\hat{\varphi}}$, with $\pmb{\hat{\varphi}}=[\hat{\varphi}_\uparrow,\hat{\varphi}_\downarrow]^{T}$, to the bosonic wave-function and evaluate the corresponding integral. By using normalization condition $|\hat{\varphi}_\uparrow|^2+|\hat{\varphi}_\downarrow|^2=N$, we derive the quantum Langevin equations \cite{Ref48,Ref49} to incorporate associated damping and noises with the cavity-atom system.
\begin{center}
	\begin{eqnarray}\label{2}
		\frac{d\hat{c}}{dt}&=&\dot{\hat c}=(i\Delta-iG \pmb{\hat{\varphi}^\dag\hat{\varphi}}-\kappa)\hat{c}+\eta+\sqrt{2\kappa} \hat c_{in},\\
		\frac{d\pmb{\hat{\varphi}}}{dt} &=&\dot{\pmb{\hat{\varphi}}} =(\frac{\Omega\sigma_{0}}{2}+\tilde{\alpha}\pmb{k_x}\sigma_{y}+
		\frac{\delta}{2}\sigma_y+\frac{\Omega_z}{2}\sigma_z-\gamma+G\hat{c}^{\dag}\hat{c})\pmb{\hat{\varphi}}\nonumber\\
		&&+\frac{1}{2}U\pmb{\hat{\varphi}}^\dag\pmb{\hat{\varphi}}\pmb{\hat{\varphi}}
		+\frac{1}{2}\varepsilon U \hat{\varphi}_\sigma^\dag\hat{\varphi}_{\acute{\sigma}}\hat{\varphi}_\sigma+\sqrt{\gamma}f_a.
	\end{eqnarray}
\end{center}
Here $\sigma,\acute{\sigma}\in\{\uparrow,\downarrow\}$ and $\kappa$ corresponds to the intra-cavity photonic decay rate. 
$\tilde{\Delta}=\Delta _{c}-NU_{0}/2$ is the effective atom-cavity detuning and $\hat{c}_{\mathrm{in}}$ represents associated Markovian 
input noise associated with cavity mode, with zero-average $\langle \hat{c}_{in}(t)\rangle=0$ over delta-correlation $\langle \hat{c}_{in}(t)\hat{c}_{in}^{\dagger}(\acute{t})\rangle=\delta(t-\acute{t})$ defined over the condition $\hbar\omega_c>>k_BT$ \cite{Dalibard2011}. 
The external harmonic trapping potential for atomic mode , which
we ignored previously as it is assumed to be spin independent, cause the atomic motional damping.
$\gamma$ corresponds to pseudo states damping and $\hat{f}_{a}$ 
is the associated Markovian noise operators with delta-correlation 
$\langle \hat{f}_{a}(t)\hat{f}_{a}^{\dagger}(\acute{t})\rangle=\delta(t-\acute{t})$ with condition $\hbar\Omega>>k_BT$. 

Further, $g_{a}=\frac{\omega_{c}}{L}\sqrt{\hbar/m_{bec}4\omega_{r}}$ is 
the atom cavity coupling, with effective recoil frequency $\Omega=4\omega_r$ and mass 
$m_{bec}=\hslash\omega_{c}^{2}/(L^{2}U^2_{0}\omega_{r})$ \cite{Ref48,Ref49}. By using the definitions of Pauli matrices, the quantum Langevin equations will read as,
\begin{widetext} 
	\begin{eqnarray}
		\frac{d\hat{c}}{dt}&=&\dot{\hat c}=(i\Delta-iG (\hat{\varphi}_\uparrow^\dag\hat{\varphi}_\uparrow+\hat{\varphi}_\downarrow^\dag\hat{\varphi}_\downarrow)-\kappa)\hat{c}+\eta+\sqrt{2\kappa} \hat c_{in}\\
		\frac{d}{dt}\binom{\hat{\varphi}_\uparrow}{\hat{\varphi}_\downarrow}&=&\begin{pmatrix}
			\frac{\Omega}{2}+\frac{\Omega_z}{2}+G\hat c^{\dag}\hat c+\frac{1}{2}UN-\gamma+\sqrt{\gamma}f_a&-i(\alpha+\frac{\delta}{2})+\frac{1}{2}U(\varepsilon-1)\hat{\varphi}_\downarrow^\dag\hat{\varphi}_\uparrow\\ 
			i(\alpha+\frac{\delta}{2})+\frac{1}{2}U(\varepsilon-1)\hat{\varphi}_\uparrow^\dag\hat{\varphi}_\downarrow&
			\frac{\Omega}{2}-\frac{\Omega_z}{2}+G\hat c^{\dag}\hat c+\frac{1}{2}UN-\gamma+\sqrt{\gamma}f_a
		\end{pmatrix}\binom{\hat{\varphi}_\uparrow}{\hat{\varphi}_\downarrow},\label{10}
		\end{eqnarray}
\end{widetext}
where, $\alpha=\tilde{\alpha}\pmb{k_x}$. The quantum Langevin equations mention above are responsible for governing the dynamics of collective system and are crucially important. 
\begin{figure}[htp]
	\includegraphics[width=4.0cm]{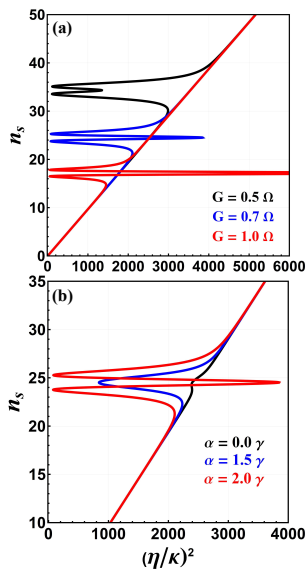}
	\caption{The steady-state behavior of cavity photon number $n_s$ as a function of normalized external pump laser intensity $\eta/\kappa$ for different cavity-atom couplings $G$ (a) and spin-orbit coupling strengths $\alpha$ (b). In (a), the black, blue, and red curves correspond $G=0.5\Omega$, $0.7\Omega$ and $1.0\Omega$, respectively, at $\alpha=2.0\gamma$. While in (b), the black, blue, and red curves correspond $\alpha=0.0\gamma$, $1.5\gamma$ and $2.0\gamma$, respectively, at $G=0.7\Omega$. The pseudo-spin states are considered with frequency $\Omega\approx 19\times 2\pi$kHz. While the other parametric are considered as $\Omega_z=0.5\Omega$, $\delta=0$, $U/\Omega=5.5$, $\Delta=2.5\kappa$, and $\kappa\approx1.3\times2\pi$MHz \cite{Refnew2,Refnew1}.}
	\label{fig2}
\end{figure}
\begin{figure*}[htp]
	\includegraphics[width=7cm]{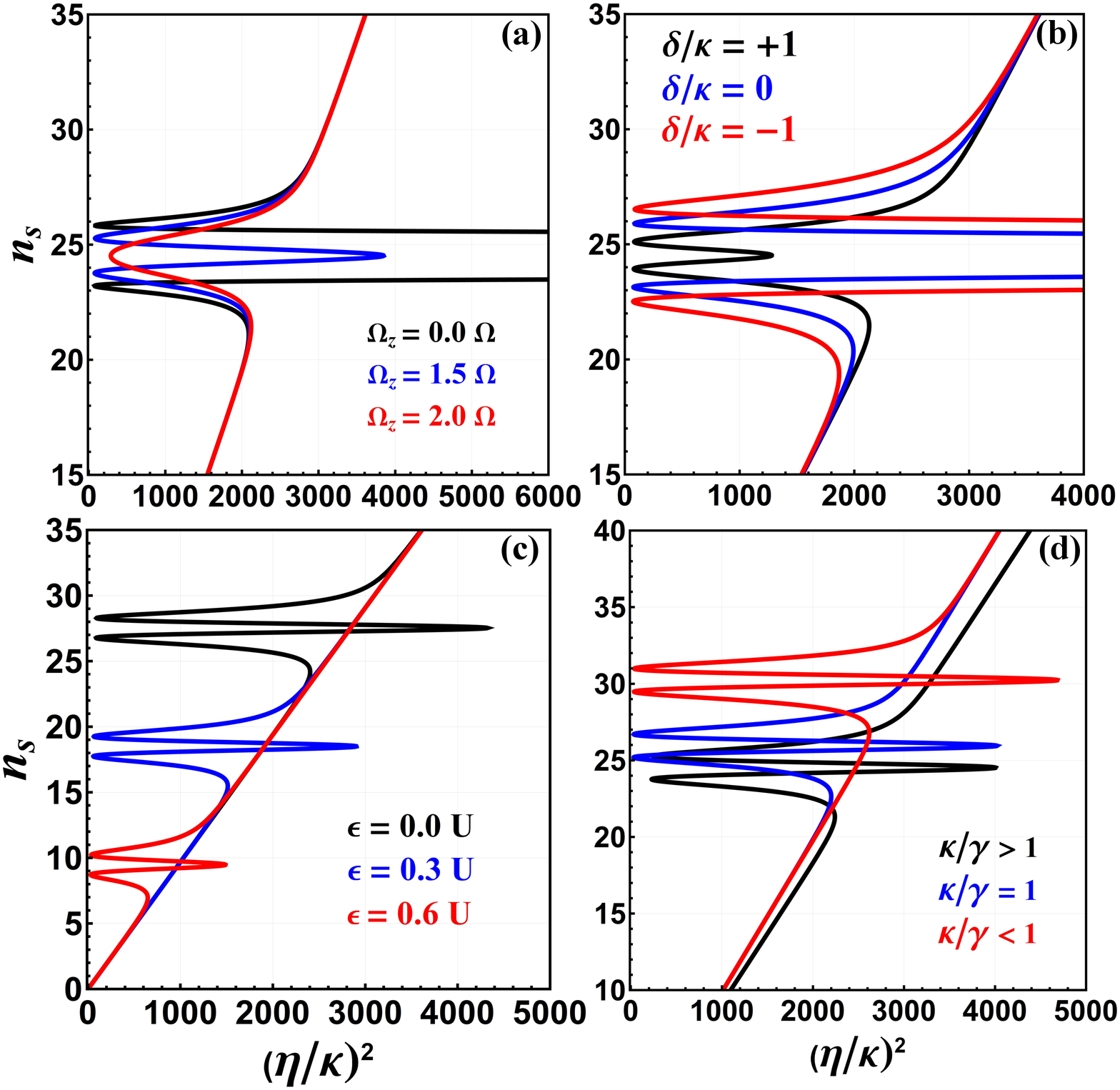}
	\caption{The steady-state cavity photon number $n_s$ as a function of $\eta/\kappa$ at constant $G=0.7\Omega$ and $\alpha=2.0\gamma$. Here (a) contains the effects of Raman coupling $\Omega_z$, while (b), (c) and (d) illustrate the influences of Raman detuning $\delta$, inter-species interaction strength $\epsilon$, and cavity decay rate $\kappa$, respectively. The other parameters used in calculations are same as in Fig.\ref{fig2}.}
	\label{fig3}
\end{figure*}

Steady-state behavior of any complex and hybrid system is crucially important in order to develop the whole picture. In this study, we are going to govern and discuss the steady-state dynamics of each associated sub-system in our setup. The steady-state behavior is obtained by assuming subsystems as classical variables (that is why we omit the motion of hat from variables) and by putting time-derivative equals to zero in Quantum Langevin equations, which will then read as, 
\begin{eqnarray}
	c_{s}&=&\frac{\eta}{\kappa +i\big(\Delta+G(\varphi_\uparrow^\dag\varphi_\uparrow+\varphi_\downarrow^\dag\varphi_\downarrow)\big)},\\
	\varphi_{\uparrow,\downarrow}&=&\frac{(\pm\alpha\mp\frac{\delta}{2})\varphi_{\downarrow,\uparrow}}{\frac{\Omega\pm i\Omega_z}{2}+Gc_s^{\dag}c_s+\frac{1}{4}UN(\varepsilon+1)-\gamma}, \\
\varphi_{\uparrow,\downarrow}^\dag&=&\frac{(\pm\alpha\mp\frac{\delta}{2})\varphi_{\downarrow,\uparrow}^\dag}{\frac{\Omega\mp i\Omega_z}{2}+Gc_s^{\dag}c_s+\frac{1}{4}UN(\varepsilon+1)-\gamma}\label{equqm}.
\end{eqnarray}\label{equ2}
The total number of steady-state photons inside the cavity can be defined as $|c_s^{\dag}c_s|=n_s$. Further, by assuming the total number bosonic particles in atomic pseudo-spin state $\uparrow$ and $\downarrow$ as $|\varphi_\uparrow^\dag\varphi_\uparrow|=N_\uparrow$ and $|\varphi_\downarrow^\dag\varphi_\downarrow|=N_\downarrow$, respectively, one can rewrite the steady-state equations as,
 \begin{eqnarray}
 	n_s&=&\frac{\eta^2}{\kappa^2 +\big(\Delta+G(N_\uparrow+N_\downarrow)\big)^2},\\
 	N_{\uparrow}&=&\frac{(\alpha-\frac{\delta}{2})^2N_{\downarrow}}{\big(\frac{\Omega}{2}+G n_s+\frac{1}{4}UN(\varepsilon+1)-\gamma\big)^2+\big(\frac{\Omega_z}{2}\big)^2},\label{11} \\
 	N_{\downarrow}&=&\frac{(\alpha-\frac{\delta}{2})^2N_{\uparrow}}{\big(\frac{\Omega}{2}+G n_s+\frac{1}{4}UN(\varepsilon+1)-\gamma\big)^2+\big(\frac{\Omega_z}{2}\big)^2}\label{22}.
 \end{eqnarray}\label{equ3}
Here, one can note that if we solve Eqs. \ref{11} and \ref{22}, then we will get a similar steady-state response for atomic number in both pseudo-spin states. The reason is the same coupling parameter between atoms and cavity because both atomic states are interacting with same cavity mode. If we consider different strengths of cavity-atom coupling, then obviously response will be different. 

\section{Multi-Stablity of Intra-Cavity Photon Number}\label{sec2}
In absence of SO-coupling $\alpha$ and Zeeman field effects $\Omega_z$ and $\delta$, the steady-state response of the system will be almost same as in cavity atomic-optomechanical system \cite{Meystre2010,kashif3}, which will be in the form of bistable states. The bistable behavior of such systems have extensively studied in literature and that is the reason of not discussing such behavior here. But in presence of SO-coupling and Zeeman field effects, the steady-state dynamics will dramatically change because now system does not have only cavity-atom coupling $G$ but its pseudo-spin states are also SO-coupled with Zeeman field effects. As the Zeeman field induced pseudo-spin states are acting like two different species whose angular momentum is coupled with each other in presence of cavity-atom coupling, therefore, the steady-state dynamics will illustrate different behavior than previous investigations, where it is simply bistable. Here, in our case, it is appeared to be multi-stable (having more than two stable states) as illustrated in Fig.\ref{fig2}, where intra-cavity photon number is plotted versus normalized external pump field coupling $\eta/\kappa$.

One can note in Fig.\ref{fig2}(a) that the steady-state cavity photon number $n_s$ possesses multiple stable states at fixed SO-coupling $\alpha=2.0\gamma$. More precisely saying, it contains three stable and two unstable states. The two unstable states correspond to two saddle points representing the occurrence of higher order nonlinearities or chaos in the system. One can some-how say that at these unstable states, cavity photon number $n_s$ following unpredictable and chaotic behavior. The three stable states determine where the $n_s$ is predictable and provide us the parametric opportunity to operate system in a stable domain. If we increase the strength of cavity-atom coupling, then these multi-stable states are appeared to be moving towards low photon number domain and enlarging towards higher intensities of external pump drive, as can be seen by blue and red curves in Fig.\ref{fig2}(a). It means that the increase in $G$ shrinks the total number photon in the steady-state, which is obvious as in strong coupling domain more photons are used in tightly binding the cavity mode. Similarly, if we keep the cavity-atom coupling constant and change the SO-coupling, then one can observe the transformation of steady-state cavity photon number from bistable dynamics to multi-stable behavior, as can be seen in Fig.\ref{fig2}(b), where $G=0.7\Omega$ is kept constant and $\alpha$ is being changed. It can be seen that in absence of SO-coupling, photon number follows bistable curvature (see black curve in Fig.\ref{fig2}(b)). But when we apply and increase SO-coupling $\alpha$, the cavity photon number shifts from bistable to multi-stable behavior, as illustrated by blue and red curves in Fig.\ref{fig2}(b), where $\alpha=1.5\gamma$ and $\alpha=2.0\gamma$, respectively. The reason, as mentioned previously, is that the pseudo-spin states are now acting as two different entities not only coupled with each other via SO-interactions but also coupled with cavity mode. Thus, these results provide evidence for the occurrence of multi-stability in $n_s$ in presence of SO-coupled atomic states. 
\begin{figure}[bp]
	\includegraphics[width=5cm]{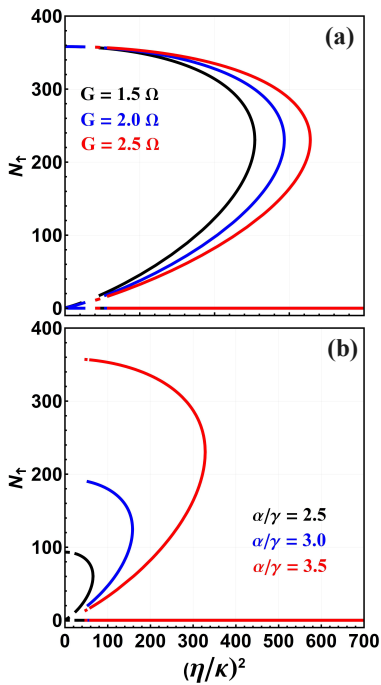}
	\caption{The steady-state atomic number of pseudo spin-$\uparrow$ state $N_\uparrow$ as a function of $\eta/\kappa$ under influence of cavity-atom couplings $G$ (a) and spin-orbit coupling strengths $\alpha$ (b). In (a), the black, blue, and red curves correspond $G=1.5\Omega$, $2.0\Omega$ and $2.5\Omega$, respectively, at $\alpha=2.0\gamma$. While in (b), the black, blue, and red curves represent $\alpha=2.5\gamma$, $3.0\gamma$ and $3.5\gamma$, respectively, at $G=2.0\Omega$. The other parameters used in calculations are same as in Fig.\ref{fig2}.}
	\label{fig4}
\end{figure}

However, these multi-stable dynamics crucially depends on the other system parameters, as illustrated in Fig.\ref{fig3}. The presence and increase in Raman coupling $\Omega_z$ is appeared to be first minimizing domain of multi-stability over the external pump intensity $\eta$ and then, at larger values, suppressing the effects of multi-stability, as illustrated by the blue and red curves of Fig.\ref{fig3}(a). It is because, at higher values of Raman coupling $\Omega_z$, the atomic pseudo states become more separated with each other yielding in more strong Zeeman field interaction \cite{Ref34}, which likewise to cavity-atom coupling results in the suppression of multi-stable photonic modes. This feature could be crucial for performing parametric optical switching in with our system. However, it also critically depends on the strength of SO-coupling. In case of Raman detuning $\delta$, the difference between photon number in different stable states is being increased with decreased in Raman detuning $\delta$, as can be seen in Fig.\ref{fig3}(b). The black, blue, and red curves represent Raman detuning $\delta=+1\kappa$, $\delta=0\kappa$ and $\delta=-1\kappa$, respectively. The decreasing in $\delta$ minimizes the gap-splitting between energy levels resulting in freeing more photons to add-up to the photonic stable states. It can also be related with asymmetric behavior of eigen energy spectrum of SO-coupled BEC \cite{Ref48}. 

Further, similarly like $\Omega_z$ and cavity-atom coupling $G$, the strength of inter-species (interaction between different pseudo-spin states) also suppresses the multi-stability photon number, as shown in Fig.\ref{fig3}(c). The reason is again the same because with higher inter-species interaction values, both pseudo-spin states more interactive with each other suppressing the number of photons in the stable states. In Fig.\ref{fig3}(d), one can observe the influence of the ratio between cavity decay rate and dissipation associated with atomic degrees fo freedom $\kappa/\gamma$. The decrease in $\kappa/\gamma$ means that the atomic dissipation is being increase as compared to cavity decay rate, as illustrated by black, blue, and red curves where $\kappa/\gamma>1$, $\kappa=1\gamma$ and $\kappa<1\gamma$, respectively. It is obvious because the atomic mechanical dissipation will directly contribute to the photonic number $n_s$ inside cavity, which will result in stable states with higher photon number. The dependence of multi-stability over all these parameters is not only crucial but also provides the controlability over steady-state dynamics \cite{kashif3}.    

\section{Multi-Stablity Analysis of Pseudo-Spin States}\label{sec3}
So-far we discussed the multi-stability of cavity photon number, but it is also very significant to investigate the steady-state behavior of atomic number (or bosonic particle number) of pseudo-spin states $N_\uparrow$ and $N_\downarrow$. In this section, we will perform the stability analysis of atomic number in pseudo-spin states. The steady-state behavior of $N_\uparrow$ and $N_\downarrow$ is appeared to different from the behavior of steady-state photon number. But, interestingly, it is also quite different from conventional atom-optomechanical systems, as illustrated in Fig.\ref{fig4}. Here, it should be noted that we just illustrated the results of pseudo spin-$\uparrow$ state because behavior spin-$\downarrow$ state will exactly be likewise. However, later we will also discuss the combined steady-state spectrum of the pseudo spin-$\uparrow$ and spin-$\downarrow$.   
\begin{figure*}[tp]
	\includegraphics[width=11cm]{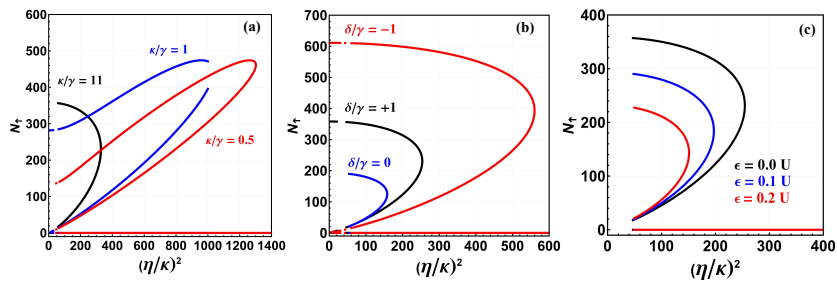}
	\caption{The steady-state population $N_\uparrow$ of atomic pseudo spin-$\uparrow$ state versus normalized $\eta/\kappa$ under the effects of cavity-decay rate $\kappa$ (a), Raman detuning $\delta$ (b) and inter-species interactions $\epsilon$ (c). In (a), the black, blue, and red curves represent $\kappa/\gamma=11$, $\kappa/\gamma=1$ and $\kappa/\gamma=0.5$, respectively. In (b), the black, blue, and red curves correspond to $\delta=+1\kappa$, $\delta=0\kappa$ and $\delta=-1\kappa$, respectively, while in (c) the black, blue, and red curves represent $\epsilon=0.0 U$, $\epsilon=0.1 U$ and $\epsilon=0.2 U$, respectively. The remaining parameters used are same as in Fig.\ref{fig2}.}
	\label{fig5}
\end{figure*}

By analyzing Fig.\ref{fig4}, one can note that the steady-state $N_\uparrow$ possesses one stable while two unstable states, which is quite unusual as compared to the previous studies. The reason behind this is the SO-interaction of pseudo-spin states of atomic degree of freedom. On one-side, these pseudo states are coupled with the cavity mode, which one the other-side they are mutually coupled with each other via SO-coupling. Because of the cavity-atom coupling $G$, the both pseudo states possess multi-stable behavior, but the stability behavior is also depending on the SO-coupling, which is generating coupling between pseudo states. It can be easily imagined by considering three balls which are mutually coupled with other. Whenever the state of one ball will be changed, it will affect the state of all balls. In previous cases, if there are multiple degrees of freedom in a cavity then they were coupled with each other via cavity field, not by directly with each other. However, we will further analyze this behavior later in this manuscript. Here, we will see the effects of different system parameter on steady-state behavior of $N_\uparrow$.

The increase in both cavity-atom coupling $G$ and SO-coupling $\alpha$ appear to be enhancing population in spin-$\uparrow$ state towards the intensity of external pump laser, as illustrated by the black, blue, and red curves of Figs.\ref{fig4}(a) and \ref{fig4}(b). However, total population $N_{\uparrow}$ is remained same. It is because both $G$ and $\alpha$ are strengthening the coupling of pseudo states with cavity mode, which makes them more dependent on the intensity of external drive. It eventually increases the mechanical effects of light on pseudo-spin states causing stable and unstable state to move further with external drive intensity, while the total population remains the same. These results could be referred as bi-unstable states because of the so-called two unstable and one stable state. Similarly, other system parameters will also alter the steady-state behavior of atomic population in spin-$\uparrow$ state, as illustrated in Fig.\ref{fig5}. The decrease in the ratio between cavity decay rate and atomic damping $\kappa/\gamma$ is yielding in strengthening the bi-unstable states towards external pump laser intensity similar to the photon number case, as illustrated in Fig.\ref{fig5}(a). The reason behind this is same that the increase in atomic damping is appearing as contribution to the atomic population. But interestingly here, the atomic population $N_\uparrow$ is initially low, with low $\kappa/\gamma$, but appeared to be increasing with increase in $\eta$. 
\begin{figure}[bp]
	\includegraphics[width=5cm]{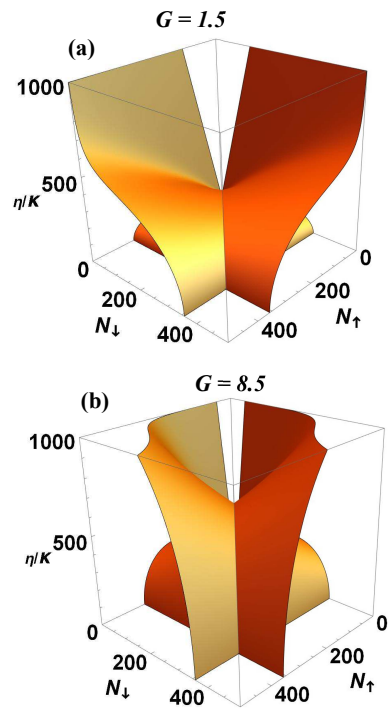}
	\caption{The combined steady-state population $N_\uparrow$ and $N_\downarrow$, atomic pseudo spin-$\uparrow$ and spin-$\downarrow$ states, respectively, versus normalized $\eta/\kappa$ with respect to the cavity-atom coupling $G$. In (a), $G=1.5\Omega$, while in (b) $G=8.5\Omega$ at $\alpha=2.0\gamma$. Here the reddish curvature indicates the stability dynamics of $N_\uparrow$, while yellowish curvature corresponds to the population $N_\downarrow$. The remaining parameters used in numerical calculation are same as in Fig.\ref{fig2}.}
	\label{fig6}
\end{figure}

However, the change in Raman detuning will dramatically alter the steady-state behavior. When the $\delta=0\kappa$, the steady-state atomic population $N_\uparrow$ possesses minimum values but when we move $\delta$ from symmetric to asymmetric case, then the population $N_\uparrow$ is significantly enhancing in bi-unstable states, as can be seen by the blue and red curves of Fig.\ref{fig5}(b), where $\delta=+\kappa$ and $\delta=-1\kappa$, respectively. The reason of it the occurrence of asymmetric shift in phase transition of atomic pseudo-spin states releasing more particle to bi-unstable states and these effects are appeared to be more prominent in negative phase shift case  $\delta=-1\kappa$. However, inter-species interaction of pseudo states is showing similar effects on bi-stable states of $N_\uparrow$ as it is showing on intra-cavity photon number state $n_s$, as illustrated in Fig.\ref{fig5}(c). The reason is also the same as more interactive pseudo-states results in less unstable particles in bi-stable atomic states. These results not only provide the evidence for the occurrence of bi-unstability but also illustrate the parametric controllability of the steady-state behavior.       
\begin{figure*}[tp]
	\includegraphics[width=10cm]{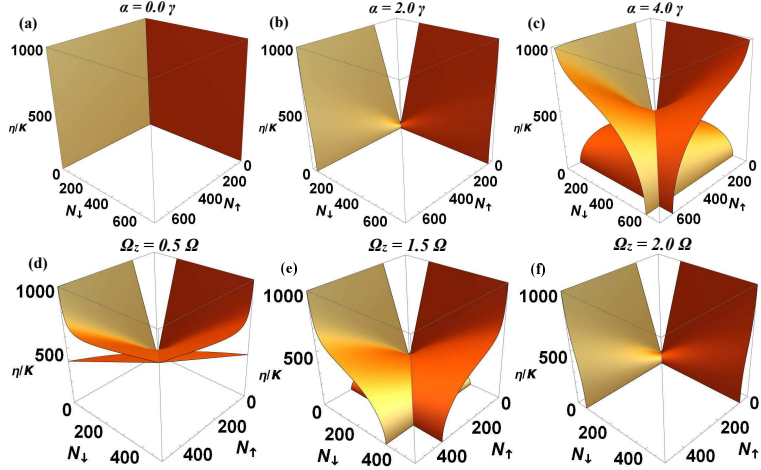}
	\caption{The collective dynamics of population $N_\uparrow$ and $N_\downarrow$ versus normalized $\eta/\kappa$ with respect to the SO-coupling $\alpha$ (a-c) and Raman coupling $\Omega_z$ (d-f). $\alpha=0.0\gamma$, $\alpha=2.0\gamma$ and $\alpha=4.0\gamma$ are using in (a), (b) and (c), respectively. While $\Omega_z=0.5\Omega$, $\Omega_z=1.5\Omega$ and $\Omega_z=2.0\Omega$ are using in (d), (e) and (f), respectively. The other parameters are same as in Fig.\ref{fig2}.}
	\label{fig7}
\end{figure*}

In order to further explore the steady-state dynamics of atomic population in different pseudo-spin states, we plot $N_\uparrow$ and $N_\downarrow$ together against normalized $\eta/\kappa$, as illustrated in Fig.\ref{fig6}. One can note the emergence of transaction between steady-state $N_\uparrow$ and $N_\downarrow$. It clearly indicates that the atomic modes are possessing phase transition between spin-$\uparrow$ and spin-$\downarrow$ states. It could also be reason of having one state and two unstable states in previous results because another stable state is appeared to be coupled with other population of other pseudo-spin state. The reason behind this is of course the SO-coupling between these sub-atomic states that is inducing another unstable state (instead of a stable state) in the pseudo state with opposite spin. In other words, the mechanical effects of light produce by the cavity mode are initially induced bistable states for atomic population but later the SO-coupling between pseudo states is converting one stable state to unstable. This type of cavity mediated SO-interaction in ultra-cold atoms are very crucial in order to bringing our new physics of multi-species hybrid interaction and has subject of various recent investigation \cite{kashif55}. If we increase the cavity-atom coupling, the transitional interface between $N_\uparrow$ and $N_\downarrow$ is appeared to be moving up at higher external pump laser intensities, as illustrated in Figs.\ref{fig6}(a) and \ref{fig6}(b), where $G=1.5\Omega$ is moved to $G=8.5\Omega$. It is obvious because the strength of cavity mode, causing the coupling between cavity and atoms, is directly proportional to the intensity of external pump. However, these all features crucially depends on other associated system parameter, which we are going discuss next in the manuscript. 

As the strength of SO-coupling $\alpha$ and Zeeman field effect induced Raman coupling $\Omega_z$ are both responsible for the splitting and creation of atomic pseudo-spin states, which are main reason for the alterations appearing in steady-state dynamics of both cavity mode and atomic pseudo states. Therefore, it is important to see the collective behavior of these pseudo spin spin-$\uparrow$ and spin-$\downarrow$ states under SO-coupling $\alpha$ and Raman coupling $\Omega_z$, as illustrated in Figs.\ref{fig7}(a-c) and \ref{fig7}(d-f), respectively. One can easily observe that, in the absence of SO-coupling $\alpha=0$, the population in both state $N_\uparrow$ and $N_\downarrow$ illustrate similar behavior and remain zero with respect to external pump intensity, see Fig.\ref{fig7}(a). It is obvious because in this configuration, there are no pseudo-spin states and BEC is acting as a single entity. However, when we start applying and increasing the SO-coupling $\alpha$, the intersection between $N_\uparrow$ and $N_\downarrow$ versus $\eta$ starts appearing, similarly as illustrated in Fig.\ref{fig6}. At lower strengths of $\alpha$, that transitional connection is occurring at lower population values. However, when we increase $\alpha$, it is shifted to higher values of $N_\uparrow$ and $N_\downarrow$, as illustrated in Figs.\ref{fig7}(b) and \ref{fig7}(c), where $\alpha=2.0\gamma$ and $\alpha=4.0\gamma$, respectively. The justification is same as given in one-dimensional case that at higher values of $\alpha$, the stable states saturate to higher population because now more atoms are interacting with not only $\alpha$ but also with $G$. In other words, at higher $\alpha$ more atoms are vulnerable to the mechanical effects of light. Similar to the one-dimensional case, the Raman coupling $\Omega_z$ here is appeared to be suppressing the number of atoms in pseudo-spin states, as illustrated in Figs.\ref{fig7}(d-f), where $\Omega_z=0.5\Omega$, $\Omega_z=1.5\Omega$ and $\Omega_z=2.0\Omega$, respectively. Not only  $N_\uparrow$ and $N_\downarrow$ are appearing to be saturating at lower values but the intersection between $N_\uparrow$ and $N_\downarrow$ is also being transferred to low atomic number domain. The reason, as mentioned earlier, is the higher order splitting between pseudo spin-$\uparrow$ and spin-$\downarrow$ state with higher Zeeman field effects resulting in less atomic bound among pseudo states. In other words, at higher Zeeman field effects, the SO-interaction dominates the mechanical effects of light and that is why less atomic population will be influenced by the external pump laser intensity.     
\begin{figure*}[tp]
	\includegraphics[width=10cm]{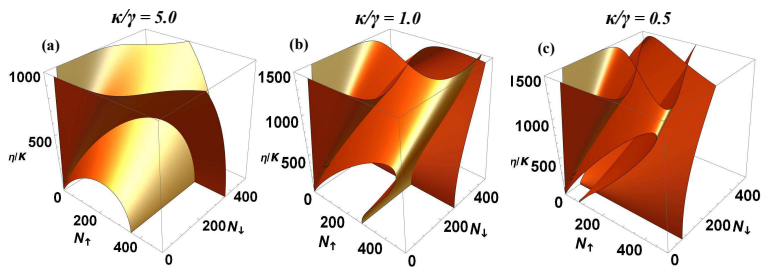}
	\caption{The collective dynamics of $N_\uparrow$ and $N_\downarrow$ with respect to normalized $\eta/\kappa$ under influence of the ratio between cavity decay rate and atomic damping $\kappa/\gamma$. (a-c) correspond to the ratio of cavity decay rate and atomic damping $\kappa/\gamma=5.0$, $\kappa/\gamma=1.0$ and $\kappa/\gamma=0.5$, respectively. The remaining parameters are same as in Fig.\ref{fig2}.}
	\label{fig8}
\end{figure*}

Moving further, the ratio between cavity decay rate and atomic damping $\kappa/\gamma$ also dramatically alters the steady-state atomic number pseudo spin-$\uparrow$ and spin-$\downarrow$ states, as shown in Figs.\ref{fig8}(a-c), where $\kappa/\gamma=5.0$, $\kappa/\gamma=1.0$ and $\kappa/\gamma=0.5$, respectively. One can note that when we increase the value of atomic dissipation $\gamma$ while keeping the cavity decay rate $\kappa$ constant, then $\gamma$ is appeared to be contributing to interactions between atomic spin-$\uparrow$ and spin-$\downarrow$ states. It results in the appears of another transitional interface between atomic population $N_\uparrow$ and $N_\downarrow$, which is indeed very interesting. It is similar to the appearance of amplification of probe light transparency from SO-coupled BEC case, see reference \cite{kashif55}, where atomic dissipation is appearing as gain to probe light. This phenomenon also led to the creation of topological edge state in transmitted bulk mode of probe light, as discussed in same study \cite{kashif55}. Thus, the appearance of secondary transitional interface between the atomic population $N_\uparrow$ and $N_\downarrow$ could be because (or cause) of the occurrence of non-trivial phase transitions of pseudo atomic spin states, and it could provide the foundation for multiple studies in this direction.

\section{Conclusion}\label{sec4}
In conclusion, we demonstrate multi-stable steady-state behavior of a cavity with spin-orbit coupled Bose-Einstein condensate. After governing steady-state dynamics, we illustrate multi-stability, with multiple stable and unstable states, of cavity photon number controllable with system parameters. Such controllable optical switching between stable and unstable states is highly crucial for modern optical devices. Further, we discuss the occurrence of similar multi-stability for the atomic population in each pseudo-spin state and show that atomic population possesses two unstable (so-called bi-unstable) states corresponding to external pump drive. The population of spin states is appeared to be also depending on each other because of the emerging transitional interface between both number when we observe them collectively. That unique feature becomes further crucial when another interface appears between spin state population by increasing atomic mechanical damping rate higher than the cavity decay rate. This could correspond to the non-trivial optical interactions of SO-coupled BEC inside cavity \cite{kashif55}. The study not only provides another direction to talk about optical switching between multiple stable and unstable states, but it is also crucial for steady-states interaction between different atomic synthetic states.

\begin{acknowledgments}
	K.A.Y. acknowledges the support of Research Fund for International Young Scientists by NSFC under grant No. KYZ04Y22050, Zhejiang Normal University research funding under grant No. ZC304021914 and Zhejiang province postdoctoral research project under grant number ZC304021952. G.X.L. acknowledges the support of National Natural Science Foundation of China under Grant Nos. 11835011 and 11774316.
\end{acknowledgments}


\begin{thebibliography}{99}
	\bibitem{Ref1}H. Walther, B. T. H. Varcoe, B. G. Englert, and T. Becker, \textit{Rep. Prog. Phys.} \textbf{69} 1325–1382 (2006).
	\bibitem{Ref2}S. Haroche and J.-M. Raimond, Exploring the Quantum: Atoms, Cavities and Photons, Oxford University Press (2006).
	\bibitem{Ref3}S. Haroche, and J. M. Raimond, \textit{Scientific American} \textbf{268} 54-60 (1993).
	\bibitem{Ref4}S. Haroche, \textit{Rev. Mod. Phys.} \textbf{85} 1083 (2013).
	
	\bibitem{Ref5}A. Blais, Arne L. Grimsmo, S. M. Girvin, and A. Wallraff, \textit{Rev. Mod. Phys.} \textbf{93}, 025005 (2021).
	\bibitem{Ref6}R. Miller \textit{et al}., \textit{J. Phys. B: At. Mol. Opt. Phys.} \textbf{38} S551 (2005).
	\bibitem{Ref7}P. Goy, J. M. Raimond, M. Gross, and S. Haroche, \textit{Phys. Rev. Lett.} \textbf{50} 1903 (1983).
	
	\bibitem{Ref8}C. Monroe, \textit{Nature} \textbf{416} 238-246 (2002).
	\bibitem{Ref9}J. Stajic, \textit{Science} \textbf{339} 1163 (2013).
		
	\bibitem{Esslinger}F. Brennecke, S. Ritter, T. Donner, and T. Esslinger, \textit{Science} \textbf{322}, 235 (2008).
	\bibitem{Meystre2010}K. Zhang, W. Chen, M. Bhattacharya, and P. Meystre, \textit{Phys. Rev. A} \textbf{81}, 013802 (2010).
	\bibitem{kashif1}K. A. Yasir, M. Ayub, and F. Saif, \textit{J. Mod. Opt.} \textbf{61}, 1318 (2014).
	\bibitem{kashif2}M. Ayub, K. A. Yasir, and F. Saif, \textit{Laser Phys.} \textbf{24}, 115503 (2014).
	\bibitem{peter2}M. Paternostro, G. D. Chiara, and G. M. Palma, \textit{Phys. Rev. Lett.} \textbf{104}, 243602 (2010).
	\bibitem{Vitali2012}M. Abdi, \textit{et al.}, \textit{Phys. Rev. Lett.} \textbf{109}, 143601 (2012).
	\bibitem{Vitali2014}M.  Abdi, \textit{et al.}, \textit{Phys. Rev. A} \textbf{89}, 022331 (2014).
	\bibitem{Sete2014}E. A. Sete, \textit{et al.}, \textit{J. Opt. Soc. Am. B} \textbf{31}, 2821 (2014).
	\bibitem{Hofer}S. G. Hofer, \textit{et al.}, \textit{Phys. Rev. A} \textbf{84}, 052327 (2011).
	\bibitem{YingPRL2012} Ying-Dan Wang and A. A. Clerk, \textit{Phys. Rev. Lett.} \textbf{108}, 153603 (2012).
	\bibitem{SinghPRL2012}S. Singh, \textit{et al.}, \textit{Phys. Rev. A} \textbf{86}, 021801  (2012).
	\bibitem{agarwal2010}G. S. Agarwal, and S. Huang, \textit{Phys. Rev. A} \textbf{81}, 041803(R) (2010).
	\bibitem{Stefan2010}S. Weis, \textit{et al.}, \textit{Science} \textbf{330}, 1520 (2010).
	\bibitem{Peng2014}B. Peng, \textit{et al.}, \textit{Nat. Commun.} \textbf{10}, 1038 (2014).
	\bibitem{SafaviNaeini2011} A. H. Safavi-Naeini, \textit{et al.}, \textit{Nature} \textbf{472}, 69 (2011).
	\bibitem{kashif11}K. A. Yasir, Zhaoxin Liang, G. Xianlong, and W. M. Liu, \textit{Eur. Phys. J. Plus} \textbf{138}, 29 (2023). 
	\bibitem{kashif111}A. Munir, K. A. Yasir, W. M. Liu, and G. Xianlong, \textit{Eur. Phys. J. Plus} \textbf{137}, 1143 (2022). 
	\bibitem{kashif3}K. A. Yasir, and W. M. Liu, \textit{Sci. Rep.} \textbf{5}, 10612 (2015).
	\bibitem{kashif33}K. A. Yasir, \textit{Opt. Commun.} \textbf{488}, 126820  (2021).
	\bibitem{kashif4}K. A. Yasir, and W. M. Liu, \textit{Sci. Rep.} \textbf{6}, 22651 (2016).
	
	\bibitem{Ref32}V. Galitski, and I. B. Spielman,  \textit{Nature} \textbf{494}, 49 (2013).
	\bibitem{Ref33}X. J. Liu \textit{et al}., \textit{Phys. Rev. Lett.} \textbf{102}, 046402 (2009).
	\bibitem{Ref34}Y. J. Lin \textit{et al}., \textit{Nature} \textbf{462}, 628 (2009).
	\bibitem{Ref35}Y. K. Kato \textit{et al}., \textit{Science} \textbf{306}, 1910-1913 (2004).
	\bibitem{Ref36}M. Konig \textit{et al}., \textit{Science} \textbf{318}, 766-770 (2007).
	\bibitem{Ref37}B. A. Bernevig, T. L. Hughes, and S. C. Zhang, \textit{Science} \textbf{314}, 1757-1761 (2006).
	\bibitem{Ref38}D. Hsieh, \textit{et al}., \textit{Nature} \textbf{452}, 970-974 (2008).
	\bibitem{Ref39}X. Li, E. Zhao, and W. V. Liu, \textit{Nat. Commun.} \textbf{4}, 4023 (2013).
	\bibitem{Ref40}K. Sun \textit{et al}., \textit{Nat. Phys.} \textbf{8}, 67-70 (2012).
	\bibitem{Ref41}J. D. Koralek \textit{et al}., \textit{Nature} \textbf{458}, 610-613 (2009).
	\bibitem{Ref42}C. Hamner, \textit{Phys. Rev. Lett.} \textbf{114}, 070401 (2015).
	\bibitem{Ref43}Y. J. Lin, K. Jim$\acute{e}$nez-Garc$\acute{\iota}$a, and I. B. Spielman, \textit{Nature} \textbf{471}, 83 (2011).
	\bibitem{Ref44}H. Hu \textit{et al}., \textit{Phys. Rev. Lett.} \textbf{108}, 010402 (2012).
	\bibitem{Ref45}Y. Deng \textit{et al}., \textit{Phys. Rev. Lett.} \textbf{112}, 143007 (2014).
	\bibitem{Ref46}B. Padhi, and S. Ghosh, \textit{Phys. Rev. A} \textbf{90}, 023627 (2014).
	\bibitem{Ref47}L. Dong \textit{et al}., \textit{Phys. Rev. A} \textbf{89}, 011602(R) (2014).
	\bibitem{Ref48}K. A. Yasir, L. Zhuang, and W. M. Liu, \textit{Phys. Rev. A} \textbf{95}, 013810 (2017).
	\bibitem{kashif55}K. A. Yasir, L. Zhuang, and W. M. Liu, \textit{npj Quantum Inf.} \textbf{8}, 109 (2022).
	
	\bibitem{Refnew0}S. Zibrov, \textit{et al}., \textit{Phys. Rev. A} \textbf{72}, 011801(R) (2005).
	\bibitem{Refnew2}F. Brennecke, \textit{et al}., \textit{Nature} \textbf{450}, 268-271 (2007).
	\bibitem{Refnew1}F. Brennecke, S. Ritter, T. Donner, and T. Esslinger, \textit{Science} \textbf{322}, 235-238 (2008).
	\bibitem{Ref49}M. Paternostro, G. D. Chiara, and G. M. Palma, \textit{Phys. Rev. Lett.} \textbf{104}, 243602 (2010).
	\bibitem{Dalibard2011}J. Dalibard \textit{et al}., \textit{Rev. Mod. Phys.} \textbf{83}, 1523 (2011).
	
	
	
	
\end{thebibliography}
\end{document}